\documentclass[conference]{IEEEtran}
\IEEEoverridecommandlockouts
\usepackage{cite}
\usepackage{amsmath,amssymb,amsfonts}
\usepackage{algorithmic}
\usepackage{graphicx}
\usepackage{textcomp}
\usepackage{xcolor}
\usepackage{subfigure}
\usepackage{booktabs}

\def\BibTeX{{\rm B\kern-.05em{\sc i\kern-.025em b}\kern-.08em
    T\kern-.1667em\lower.7ex\hbox{E}\kern-.125emX}}
\begin{document}

\title{MetaSID: Singer Identification with Domain Adaptation for Metaverse\\
\thanks{$\ast$Corresponding author: Jianzong Wang (jzwang@188.com).}
}

\author{\IEEEauthorblockN{Xulong Zhang, Jianzong Wang$^{\ast}$, Ning Cheng, Jing Xiao}
\IEEEauthorblockA{\textit{Ping An Technology (Shenzhen) Co., Ltd., China} }
}
\maketitle

\begin{abstract}
Metaverse has stretched the real world into unlimited space. There will be more live concerts in Metaverse. The task of singer identification is to identify the song belongs to which singer. However, there has been a tough problem in singer identification, which is the different live effects. The studio version is different from the live version, the data distribution of the training set and the test set are different, and the performance of the classifier decreases. This paper proposes the use of the domain adaptation method to solve the live effect in singer identification. Three methods of domain adaptation combined with Convolutional Recurrent Neural Network (CRNN) are designed, which are Maximum Mean Discrepancy (MMD), gradient reversal (Revgrad), and Contrastive Adaptation Network (CAN). MMD is a distance-based method, which adds domain loss. Revgrad is based on the idea that learned features can represent different domain samples. CAN is based on class adaptation, it takes into account the correspondence between the categories of the source domain and target domain. Experimental results on the public dataset of Artist20 show that CRNN-MMD leads to an improvement over the baseline CRNN by 0.14. The CRNN-RevGrad outperforms the baseline by 0.21. The CRNN-CAN achieved state of the art with the F1 measure value of 0.83 on album split. 
\end{abstract}

\begin{IEEEkeywords}
Metaverse, Singer identification, Domain adaptation, CNN, GRU
\end{IEEEkeywords}

\section{Introduction}

The Metaverse~\cite{lee2021all} breaks the limitations of the physical boundaries of concert venues, it can bring tens of millions of viewers to an audio-visual feast. Singer identification (SID), also known as artist classification, is an interesting task under the subject of music information retrieval (MIR)~\cite{gao2021vocal,zhang2020research}. It aims to name the singer in a given audio sample to facilitate users to find and manage the music library. If trained properly, the feature representation of the singing voices will also be learned by the SID model, which can be used in the vocal related tasks, such as similarity singer search, playlist generation, or singing synthesis~\cite{hsieh2020addressing,csmt2021sun,lee2019learning,sibo2022,chen2019practical,liu2019score,humphrey2018introduction,zhang2020singing}.

The studies on SID can be categorized into three aspects, there are feature engineering, traditional classifier, and the deep model. In~\cite{ellis2007classifying}, Ellis \textit{et al.} investigate fusion timbral and Chroma features for artist classification. Eghbal-zadeh \textit{et al.}~\cite{eghbal2015vectors} proposed an approach to extract song-level descriptors built from frame-level timbral features. The biggest limitation of feature engineering is that it is difficult to find a feature that accurately describes different singing voices~\cite{wang2020transfer}. In addition to the feature representation, most research focus on the classifier. Different classifiers have been tried on singer identification, including SVM, GMM, HMM, and random forest~\cite{zhang2021singer,sangeetha2020singer,gao2021vocal,aolan2021}. With the successful application of deep models in various tasks~\cite{zhao2022nnspeech,csmt2021sun}, some studies are using deep models to improve performance on singer identification, such as CRNN~\cite{nasrullah2019music} which is a state of the art method.

SID is a so far unsolved task, there is at least one challenge how to deal with the mismatch among the different styles of singing. The songs of existing studio songs are easy to obtain, and the accuracy of identifying the songs in the existing album can be very high. However, when the model is applied to the songs from the new style album or live performance version, it will bring a great performance degradation~\cite{zhang2019novel}. The singer singing in the live performance will have style changed slightly, and even sing other singers sing without being collected by his album. The sing styles of songs included in different albums are usually distinct, and the styles of songs in the same album are often similar. This dissimilarity is called the album effect~\cite{su2013sparse,zhang2020research} in the task of singer identification. 
\begin{figure}[htbp]
  \includegraphics[width=\linewidth]{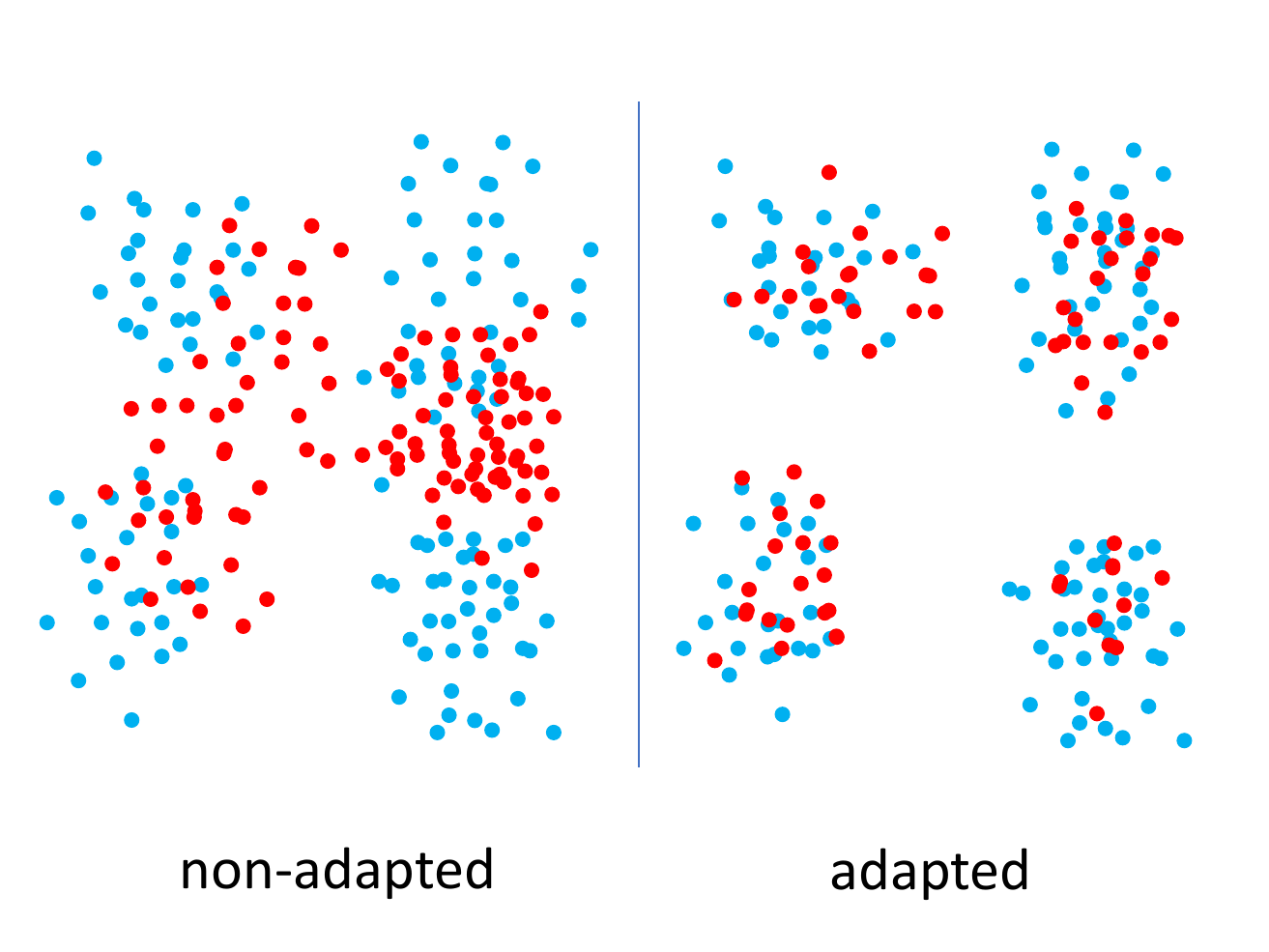}
  \caption{The schematic diagram of domain adaptation. The blue point is in the source domain, and the red point is in the target domain.}
  \label{fig:1}
\end{figure}

Domain adaptation (DA)~\cite{bruzzone2009domain, chu2013selective,asru2021zhang,gong2013connecting,pan2010domain,tang2022avqvc} utilizes supervised data in the source domain to execute similar tasks in the target domain with unsupervised data, which is a specific case of transfer learning. Figure \ref{fig:1} is a diagram showing the effect of the domain adaptation. If it is non-adapted, the data distribution of the source domain and the target domain are quite different. With DA, it is often possible to learn a common shared space that can sample the data from the source domain same as the target domain with similar distributions. There are several methods to make the distribution shift between the domains of source and target. The most commonly used methods are maximum mean discrepancy (MMD)~\cite{long2017deep, yan2017mind,zhang2022Singer,long2015learning, qubo2021,long2016unsupervised, long2016unsupervised,ghifary2014domain}, which make a comparison and reduction for the distribution shift. Such as the related work by Tzeng \textit{et al.}~\cite{tzeng2014deep}, a deep domain confusion network (DDC) was proposed. The DDC used two CNNs with shared weights for the modeling of the source domain and target domain. The whole network is used for the task of classification with the optimization of classification loss, while the MMD metric is applied as an adaptation layer for the measuring of the different domains. 

In this paper, we draw on the idea of DDC and the fusion of the structure for singer identification. Domain adaptation is fused with CRNN~\cite{choi2016explaining,zhang2022SUSing, lee2019learning}to solve the problem of the album effect. These three methods of domain adaptation were proposed for the task of singer identification. Finally, the proposed three methods are the distance-based method CRNN-MMD, the gradient reversal method CRNN-RevGrad, and the clustering distance-based CRNN-CAN.

\section{Related Works}

\subsection{Baseline Method of CRNN for SID}
\begin{figure}[htbp] 
  \centering
  \includegraphics[width=0.7\linewidth]{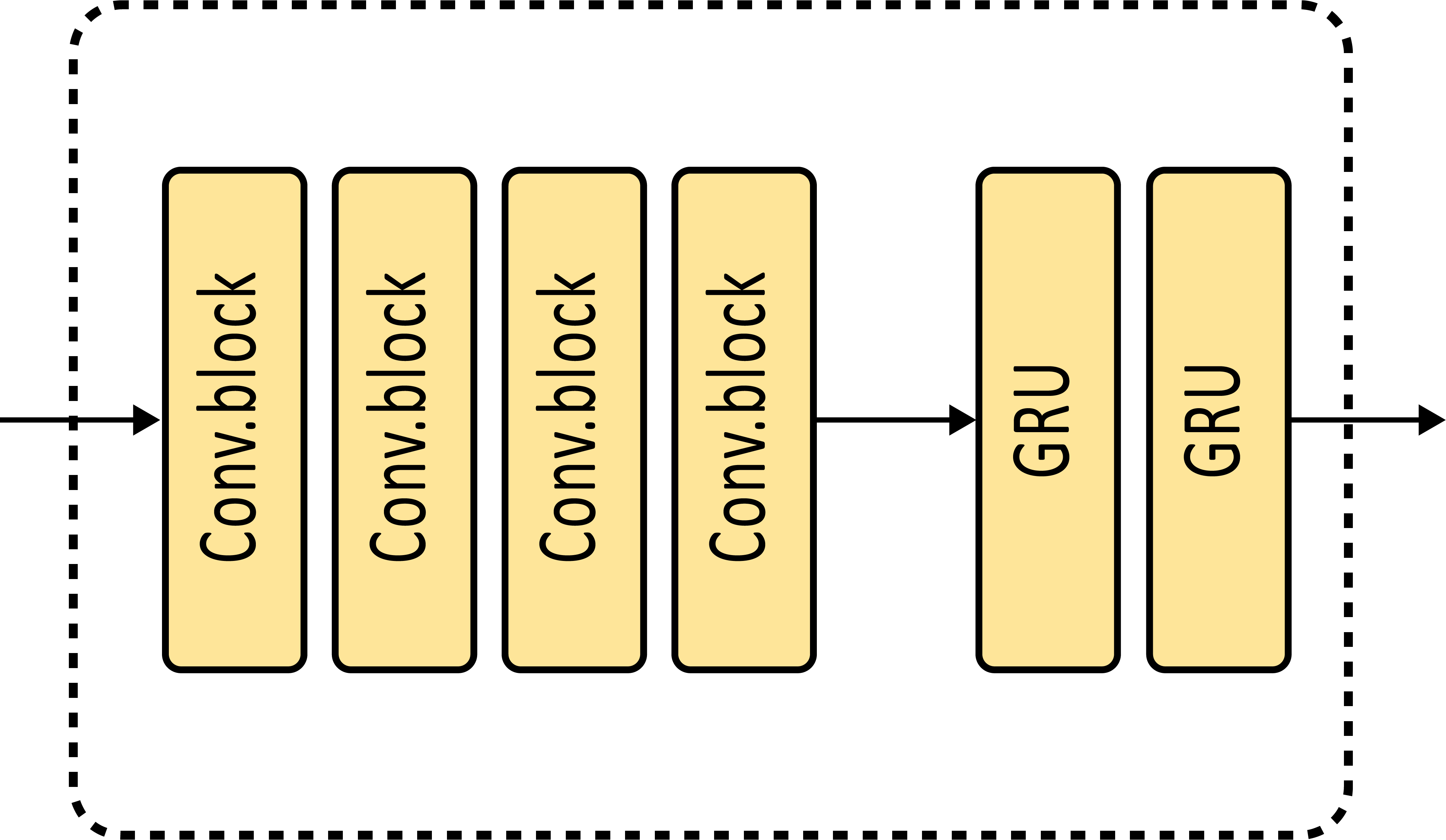}
  \caption{The model architecture of baseline method of CRNN for SID}
  \label{fig:crnn}
\end{figure}
The state-of-the-art method of SID, CRNN~\cite{nasrullah2019music}, was chosen as the baseline. As shown in Figure \ref{fig:crnn}, the model mainly consists of four CNN blocks and two GRU blocks. Then connected to a fully connected layer and a softmax layer for classification. CRNN combines the advantages of CNN and RNN. Choi \textit{et al.}~\cite{choi2017convolutional} have proved that CNN could be used as a feature extractor for MIR tasks. At the same time, due to the characteristic that the audio is continuous along the time axis, the recurrent neural network (RNN) has also achieved great success in audio tasks such as speech recognition~\cite{ellis2007classifying,wang2022drvc}. The average F1 score on the album-split Artist20 dataset is \textbf{0.603}, which is a significant improvement compared to the previous model. 

\subsection{Maximum Mean Discrepancy (MMD)}

The MMD is based on the fact that, if the distribution of the source domain is identical with the target domain, the related statistic parameters should be similar or same. In MMD, from the marginal distributions $P(X^s)$ and $Q(X^t)$ sampled ${x_i^s}$ and ${x_i^t}$ respectively, the data sampling is $i.i.d$. For the samples, MMD conduct a test of the two-sample by a kernel to determine whether satisfy the hypothesis $P=Q$ or not. Formally, MMD defines the difference between two domains with their feature vector in the reproducing kernel Hilbert space (RKHS), \textit{i.e.} 
\begin{equation}
\mathcal{D}_{\mathcal{H}}(P, Q) \triangleq \sup _{f \sim \mathcal{H}}\left(\mathbb{E}_{\boldsymbol{X}^{s}}\left[f\left(\boldsymbol{X}^{s}\right)\right]-\mathbb{E}_{\boldsymbol{X}^{t}}\left[f\left(\boldsymbol{X}^{t}\right)\right]\right)_{\mathcal{H}}
\end{equation}
where $\mathcal{H}$ refers to the set of all functions $f$ that map the feature space to the real number set $R$. In fact, the empirical kernel mean embedding estimated the MMD in form of squared value, it build the layer $l$,
\begin{equation}
\begin{aligned}
\hat{\mathcal{D}}_{l}^{m m d} &=\frac{1}{n_{s}^{2}} \sum_{i=1}^{n_{s}} \sum_{j=1}^{n_{s}} k_{l}\left(\phi_{l}\left(\boldsymbol{x}_{i}^{s}\right), \phi_{l}\left(\boldsymbol{x}_{j}^{s}\right)\right) \\
&+\frac{1}{n_{t}^{2}} \sum_{i=1}^{n_{t}} \sum_{j=1}^{n_{t}} k_{l}\left(\phi_{l}\left(\boldsymbol{x}_{i}^{t}\right), \phi_{l}\left(\boldsymbol{x}_{j}^{t}\right)\right) \\
&-\frac{2}{n_{s} n_{t}} \sum_{i=1}^{n_{s}} \sum_{j=1}^{n_{t}} k_{l}\left(\phi_{l}\left(\boldsymbol{x}_{i}^{s}\right), \phi_{l}\left(\boldsymbol{x}_{j}^{t}\right)\right)
\end{aligned}
\end{equation}
where $x^s \in S^{\prime} \in S$, $x^t \in T^{\prime} \in T$, $n_s = |S^{\prime}|$. The $S^{\prime}$ and $T^{\prime}$ are the data sampled from $S$ and $T$ within a batch size respectively. For the deep neural network, the kernel of the $l$-th layer is represented as $k_l$. 

\subsection{Gradient Reversal (RevGrad)}

The basic idea of RevGrad~\cite{ganin2016domain} is to use a similar training method with GAN to let the generator generate features, and then let the discriminator determine whether it is a feature of the source domain or the target domain. If the discriminator cannot distinguish which domain it is, it means that the source domain and target domain are consistent in the learned feature space.

Consider the classification task with input space of $X$, and $Y={0,\dots,N}$ is the possible labels of the predicted class. There are at least two domains over the data of $(X,Y)$, where we call the known label domain as source $D_s$, and the other domain as target $D_t$. Formally express as Equation~\ref{eq:da S} and Equation~\ref{eq:da T}.

\begin{equation}
\label{eq:da S}
    S=\{(x_i, y_i)\}^n_{i=1} \sim (D_s)^n
\end{equation}

\begin{equation}
    \label{eq:da T}
    T=\{x_i\}^N_{i=n+1} \sim (D_t^X)^{n^{'}}
\end{equation}
where where $D^X_t$ is the marginal distribution of
$D_t$ over $X$. With the total $n+n^{'}$ samples to learn a classifier for $X \rightarrow Y$ without information label of $D_t$.

The optimization process can use the minimization-maximization idea to train. The optimization algorithm of the model is carried out in the following three steps.
\begin{itemize}
  \item Train the source feature extractor $M_S$ and classifier $C$ on the source domain.
  \item Fix $M_S$ and $C$, train the target feature extractor $M_T$ and domain discriminator $D$.
  \item Use $M_T$ and $C$ to classify the target domain.
\end{itemize}

\subsection{Contrastive Adaptation Network (CAN)}

\begin{figure}[t]
  \includegraphics[width=\linewidth]{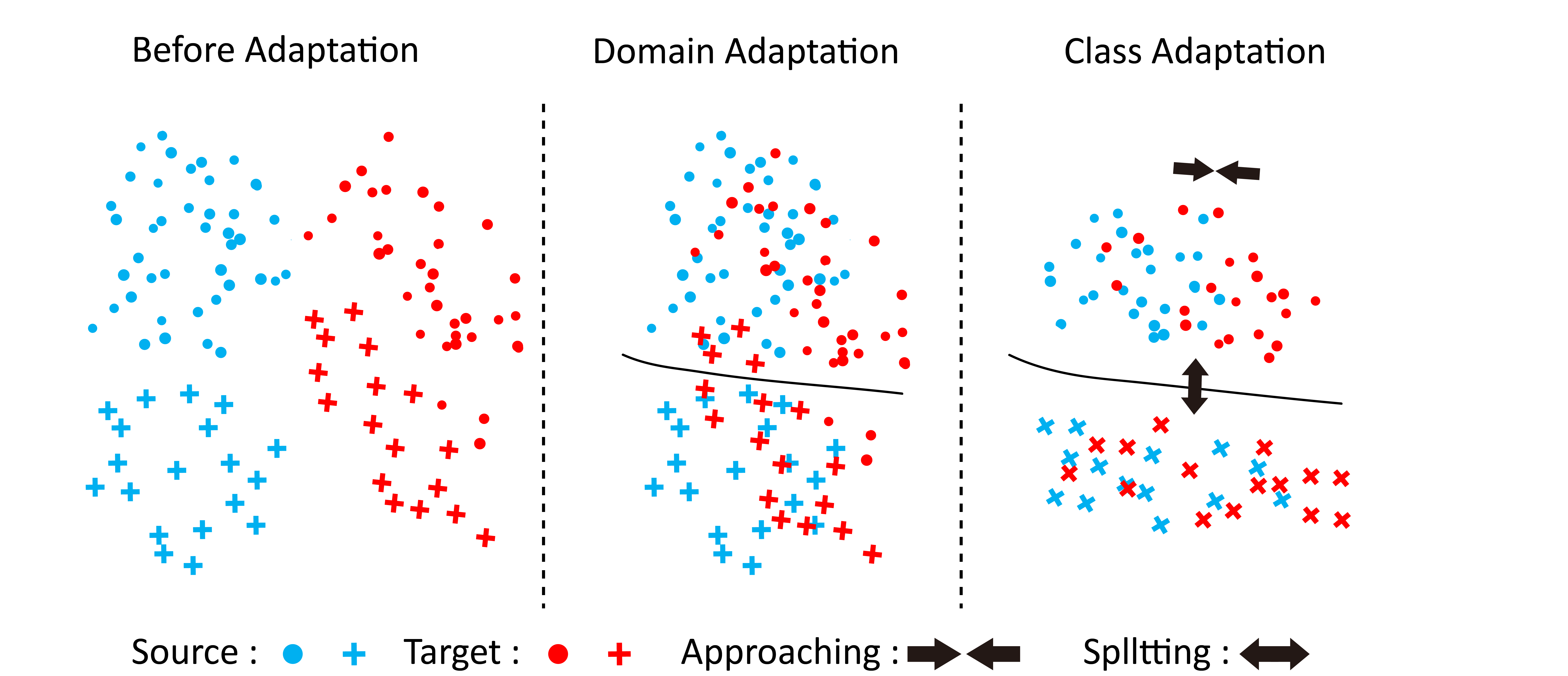}
  \caption{Comparison between previous domain-discrepancy minimization methods with class-discrepancy minimization methods}
  \label{fig:CAN-shiyitu}
\end{figure}

As shown in Figure \ref{fig:CAN-shiyitu}, the MMD and RevGrad domain adaptation methods mentioned above do not consider the correspondence between categories, which may cause the disorder of the correspondence between domains. The performance of the domain adaptation can be impaired by several problems as below. First, the classes of the source domain and target domain may be aligned not match, \textit{e.g.} MMD can be optimized even when there are different classes from the source domain dataset and the target domain dataset. Second, the learned model was classified well in the source classes but had a poor ability to generalization for the classes in the target domain. Due to the existence of many suboptimal classifications near the decision boundary, it could lead to an overfit for the source domain dataset but could not classify the target domain classes very well.

While the CAN~\cite{aljanaki2018data,zhang2022TDASS} take into account the class information of both the source domain and target domain, it also measures the same class across the domains and the different class across the domains. The Contrastive Domain Discrepancy (CDD) in CAN is established on the difference between conditional data distributions across domains. The CDD is calculated as 
\begin{equation}
\begin{split}
    D^{cdd}&=\frac{1}{M}\sum_{c=1}^{M}D^{cc}(y_{1:n_t}^t,\phi )\\
    &-\frac{1}{M(M-1)}\sum_{c=1}^{M}\sum_{{c}'=1 \& {c}'\neq c} ^{M}D^{c{c}'}(y_{1:n_t}^t,\phi ),
\end{split}
\end{equation}
where $M$ is the classes number, $y^t$ is the target data label, for the deep neural network, $\phi$ is used as the mapping from the input to a following layer. The data samples number of a batch from target domain is $n_t$. The discrepancy of intra-class across domains is  $D^{cc}$, and the discrepancy of inter-class across domains is  $D^{c{c}'}$. In the optimization, it minimize the inter-class domain discrepancies and maximize the intra-class domain discrepancies.

\section{Methodology}

\begin{figure*}[htbp]
\centering
\subfigure[CRNN-MMD]{
\includegraphics[width=0.29\linewidth]{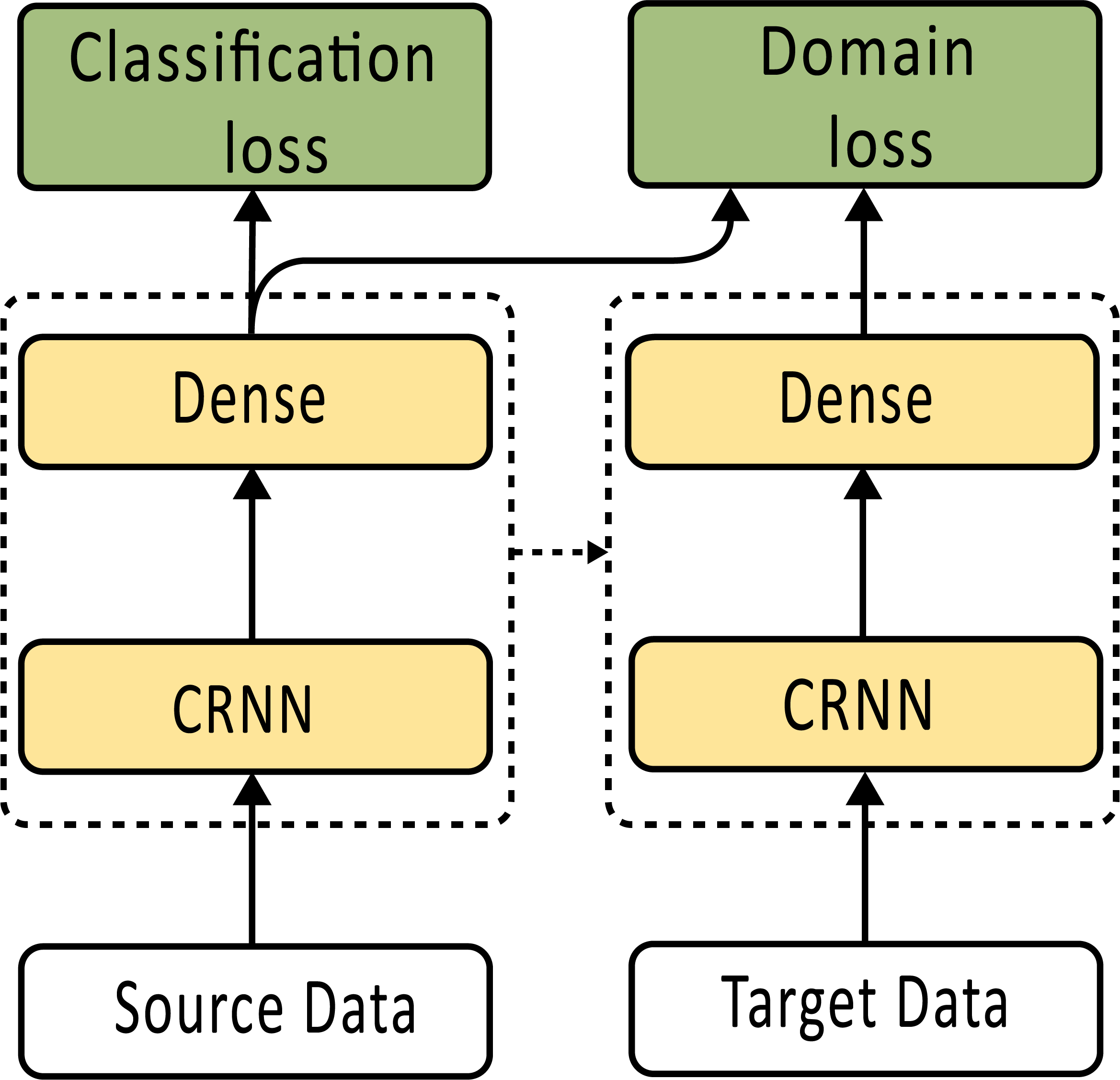}
\label{fig:architecture mmd}
}
\hspace{0mm}
\subfigure[CRNN-RevGrad]{
\includegraphics[width=0.30\linewidth]{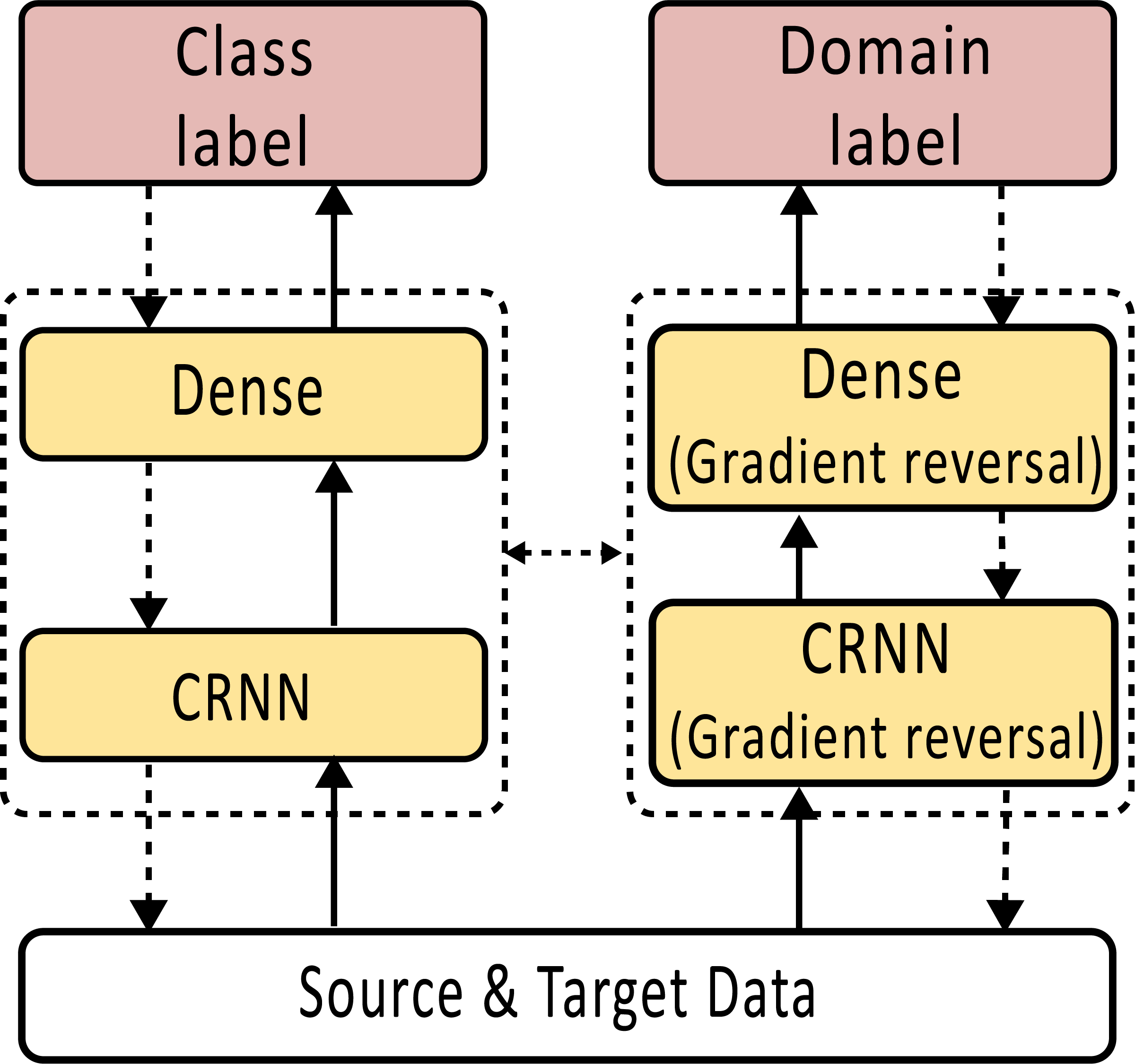}
\label{fig:architecture revgrad}
}
\hspace{0mm}
\subfigure[CRNN-CAN]{
\includegraphics[width=0.32\linewidth]{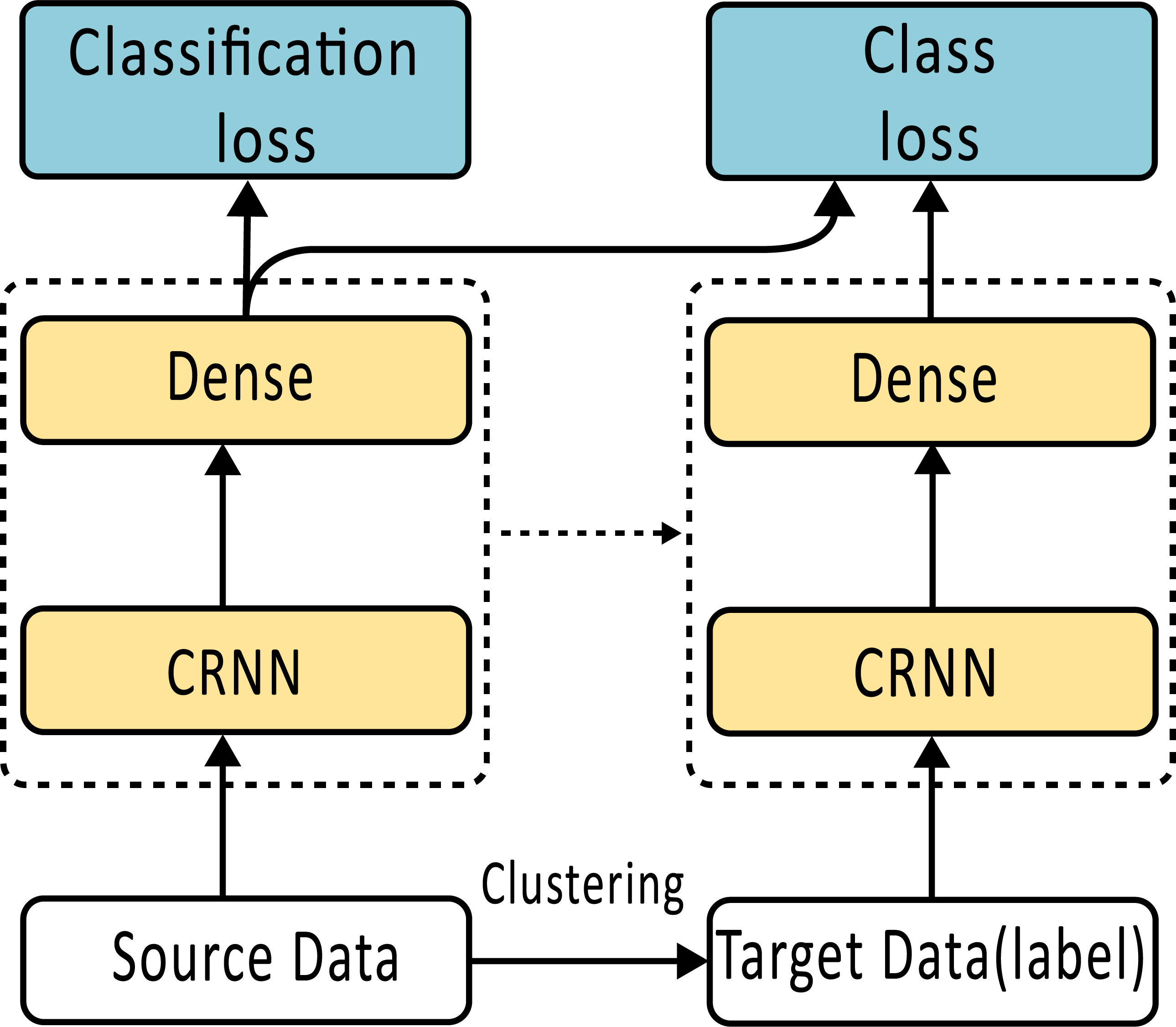}
\label{fig:architecture can}
}
\caption{The network architecture of the proposed three methods}
  \label{fig:architecture}
\end{figure*}

Combined with the deep model CRNN, we proposed three singer identification methods based on domain adaptation methods. 

\subsection{CRNN-MMD}
 Given two distributions $s$ and $t$, the MMD is defined as follows:

\begin{equation}
MMD^{2}(s, t)=\sup _{\|\phi\|_{\mathcal{H}} \leq 1}\left\|E_{\mathrm{x}^{s} \sim s}\left[\phi\left(\mathrm{x}^{s}\right)\right]-E_{\mathrm{x}^{t} \sim s}\left[\phi\left(\mathrm{x}^{t}\right)\right]\right\|_{\mathcal{H}}^{2}
\end{equation}
where $\phi$ represents the kernel function that maps the original data to a reproducing kernel Hilbert space (RKHS) and ${\|\phi\|_{\mathcal{H}} \leq 1}$ defines a set of functions in the unit ball of RKHS $\mathcal{H}$.

As shown in Figure \ref{fig:architecture mmd}, we use a single hidden layer in the feedforward neural networks for the calculation of the MMD metric. From the different domains' representation of the embedding in the latent space, to reduce the mismatch of different distributions can calculate the metric of MMD. We use a kernel two-sample test for the effective metric of the MMD. The estimation of the MMD is given as below:

\begin{equation}
MMD^{2}\left(D_{s}, D_{t}\right)=\left\|\frac{1}{M} \sum_{i=1}^{M} \phi\left(\mathrm{x}_{i}^{s}\right)-\frac{1}{N} \sum_{j=1}^{N} \phi\left(\mathrm{x}_{j}^{t}\right)\right\|^{2}
\end{equation}

\subsection{CRNN-RevGrad}

The CRNN-RevGrad shown in Figure \ref{fig:architecture revgrad} is to add a gradient reversal to the CRNN and Dense layer. The feature extractor consists of 4 CNN blocks and 2 GRU blocks same as baseline CRNN. The class label is used for the classifier that classifies the features of the source domain data. The Domain label is a discriminator that distinguishes the characteristics of the source domain data and the target domain data. This discriminator should be continuously enhanced so that it can distinguish the characteristics of the source domain or the target domain. At the same time, the extractor should be enhanced so that the extracted features can confuse the discrimination of the discriminator. In this way, the finally extracted features are consistent in the space of the source domain and the target domain. 

The model of CRNN-RevGrad to keep the feature distribution across the source domain and target domain integrates a gradient reversal into the standard architecture. The network is built up of several layers for feature extraction with shared weights and two classifiers. The optimization of CRNN-RevGrad, minimizes the domain confusion loss for all samples from the source domain and target domain, and it minimizes the classification loss with the labels from the source domain dataset, while for the target domain data samples it maximizes domain confusion loss with the use of gradient reversal.

CRNN-RevGrad minimizes the distance of the representation of source domain and target domain through the function as below:
\begin{equation}
\begin{aligned}
\min _{M^{s}, C} \mathcal{L}_{c l s}\left(X^{s}, Y^{s}\right)=\\\quad{E}_{\left(x^{s}, y^{s}\right) \sim\left(X^{s}, Y^{s}\right)}\sum_{k=1}^{K}{1}_{\left[k=y^{s}\right]} \log C\left(M^{s}\left(x^{s}\right)\right)
\end{aligned}
\end{equation}
\begin{equation}
\begin{aligned}
\min _{D} \mathcal{L}_{a d v D}\left(X^{s}, X^{t}, M^{s}, M^{t}\right)=\\
-{E}_{\left(x^{s}\right) \sim\left(X^{s}\right)}\left[\log D\left(M^{s}\left(x^{s}\right)\right)\right]\\
-{E}_{\left(x^{t}\right) \sim\left(X^{t}\right)}\left[\log \left(1-D\left(M^{t}\left(x^{t}\right)\right)\right)\right]
\end{aligned}
\end{equation}
\begin{equation}
\begin{aligned}
\min _{M^{s}, M^{t}} \mathcal{L}_{a d v M}\left(M^{s}, M^{t}\right)=\\
-{E}_{\left(x^{t}\right) \sim\left(X^{t}\right)}\left[\log D\left(M^{t}\left(x^{t}\right)\right)\right]
\end{aligned}
\end{equation}
where $M_s$ and $M_t$ is the representation of the mappings learned from the source domain samples $X_s$ and target domain samples$X_t$. The classifier $C$ is used for the source domain. Trough the supervised data from the source domain , the classifier could be trained with the classification loss function of $L_{cls}$. There are also two loss functions of $L_{adv}D$ and $L_{adv}M$. $L_{adv}D$ is used for the discriminator, and it need to be minimized. $L_{adv}M$ is to learn a domain invariant representation, and it need to be maximized.

\subsection{CRNN-CAN}
Considering the example in Figure \ref{fig:CAN-shiyitu}, the objective function of Contrastive Domain Discrepancy (CDD) of CAN is used for the unsupervised domain adaption with the class aligned across the domains. The data samples from the same class of the source domain and target domain will be drawn closer by the metric of CDD. 

Form the related studies in the artist classification, the CRNN can learn more transferable features than shallow models. We start from CRNN pre-trained networks as the backbone, the last layer of the fully connected (FC) layer was replaced with a task-specific layer. The CRNN-CAN has shown in Figure \ref{fig:architecture can}, through back-propagation to minimize the domain discrepancy, the network of the last FC layer and the CRNN module were fine-tuned.

To minimize the CDD over the deep neural network build up with several FC layers, we show the objective function as below, \textit{i.e.} minimizing
\begin{equation}
    D_{\mathcal{L}}^{cdd}=\sum _{l=1}^{L} D_l^{cdd}
\end{equation}
where $l$ is the FC layer, and $L$ is the total number of FC layers. The network was trained with labeled sources data through minimizing the cross-entropy loss,
\begin{equation}
    loss^{ce}=-\frac{1}{{n}'_s}\sum_{{i}'=1}^{{n}'_s}logP_\theta (y_{{i}'}^s|x_{{i}'})
\end{equation}
where the sample $x^s$ of source domain has the label of $y^s\in {0,1,\dots,M-1}$. The predicted probability of $y$ is show as $P_{\theta}(y|x)$, it has a weight of $\theta$, and the input is $x$.

Therefore, the overall objective can be formulated as 
\begin{equation}
    \min_\theta loss=loss^{ce}+\beta D_{\mathcal{L}}^{cdd}
\end{equation}
where the discrepancy penalty term is multiply with the weight factor of $\beta$.

To estimate and optimize with CDD, we need labels of the target domain. However, target labels are unknown. Additionally, during the training in a batch, for class C, there are only contain samples in one domain. To solve the two problems, we adapt two tricks during the training phase. We use CAN alternatively to estimate the label hypothesis of target samples through clustering. After clustering, when estimating CDD, fuzzy target data far from the cluster center is set to zero, and fuzzy classes with few target samples around the cluster center are set to zero. Moreover, to satisfy the training in the manner of mini-batch size, both source and target domains were sampled for each batch, i.e. we randomly sample data from the source domain and target domain for each class.
\section{Experiments}
\begin{figure*}[htbp]
\centering
\subfigure[source domain data]{
\includegraphics[width=0.33\linewidth]{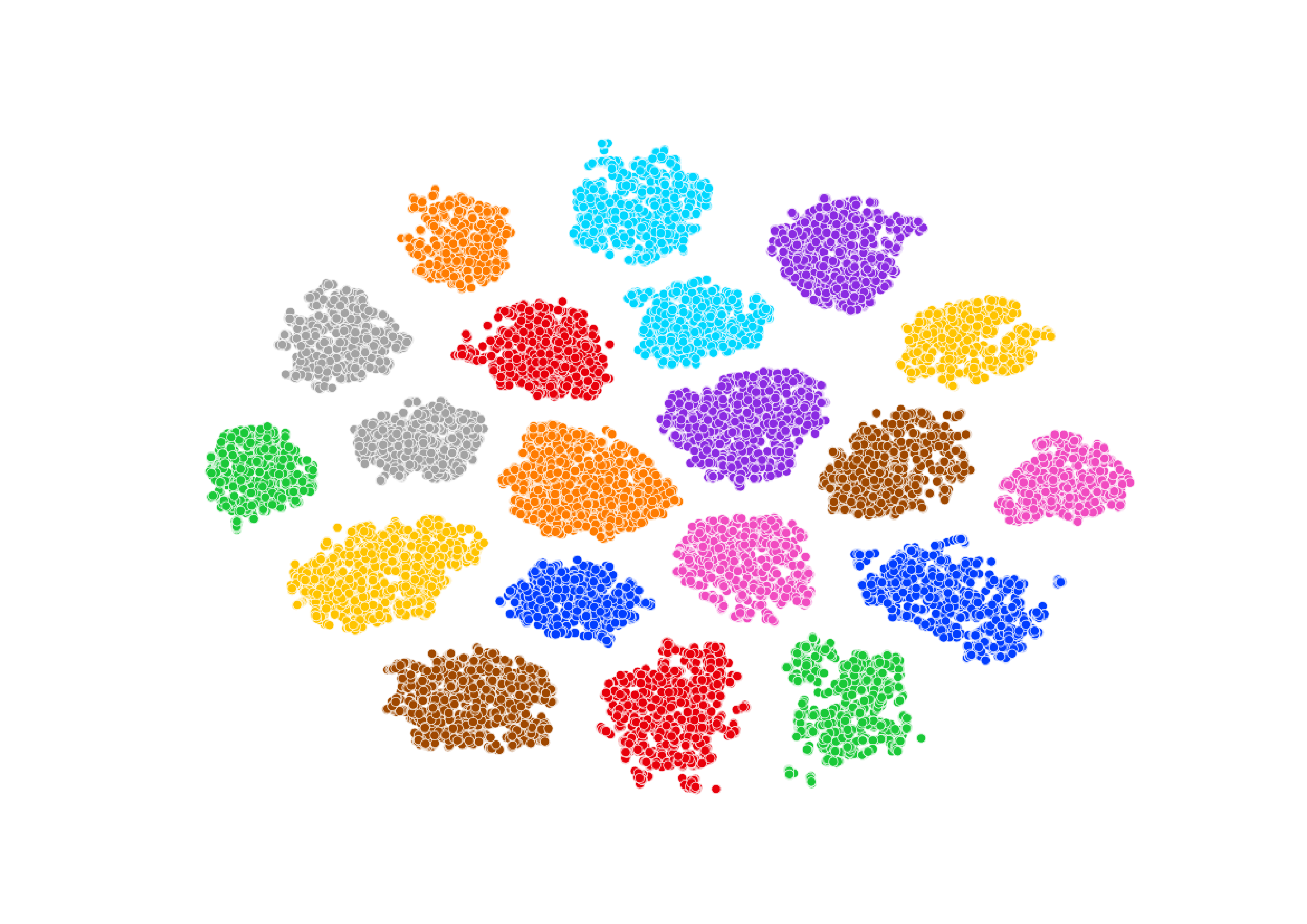}
\label{fig:source domain}
}
\hspace{-5mm}
\subfigure[target domain data]{
\includegraphics[width=0.33\linewidth]{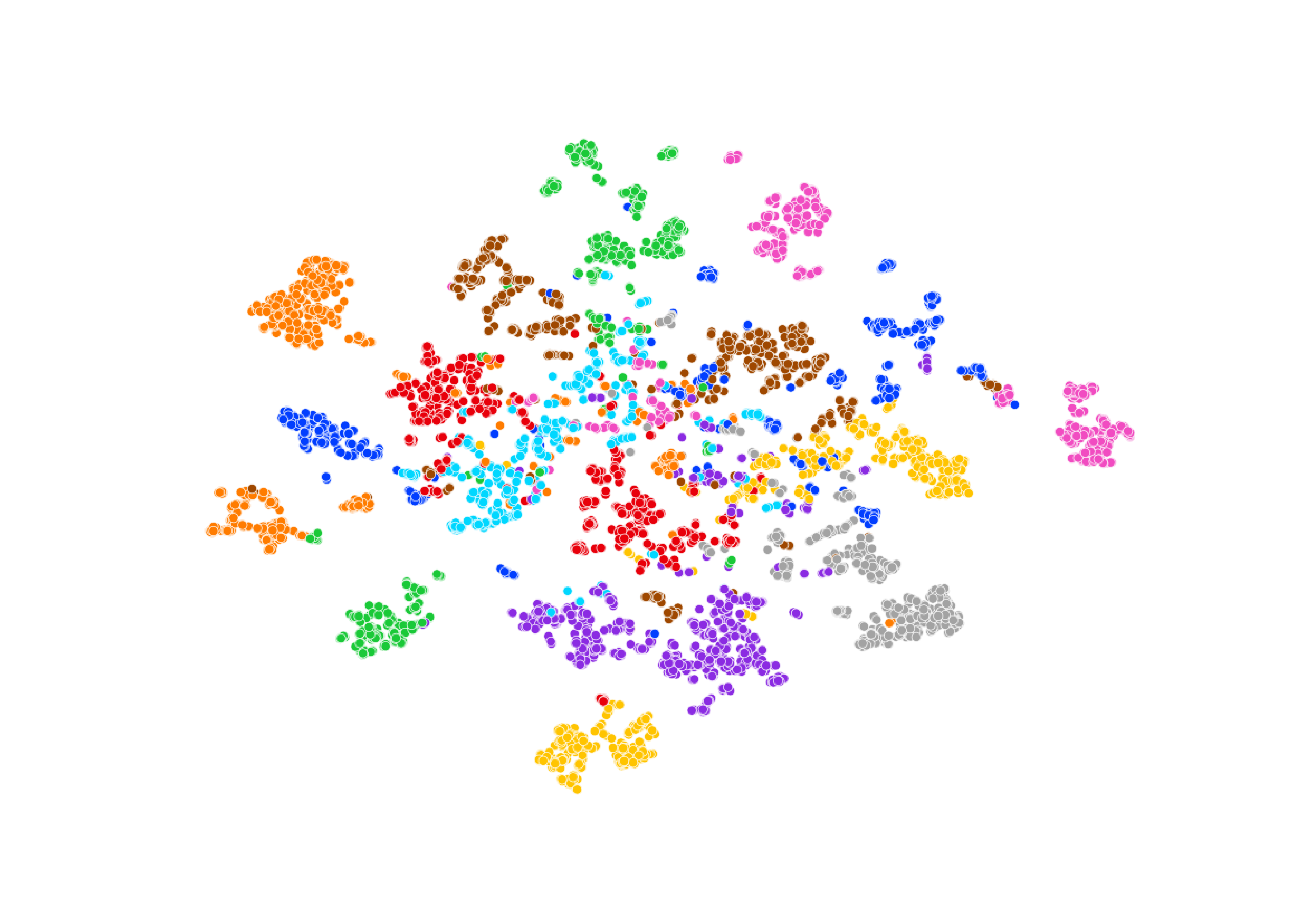}
\label{fig:target domain}
}
\hspace{-5mm}
\subfigure[feature represent with domain adaption]{
\includegraphics[width=0.33\linewidth]{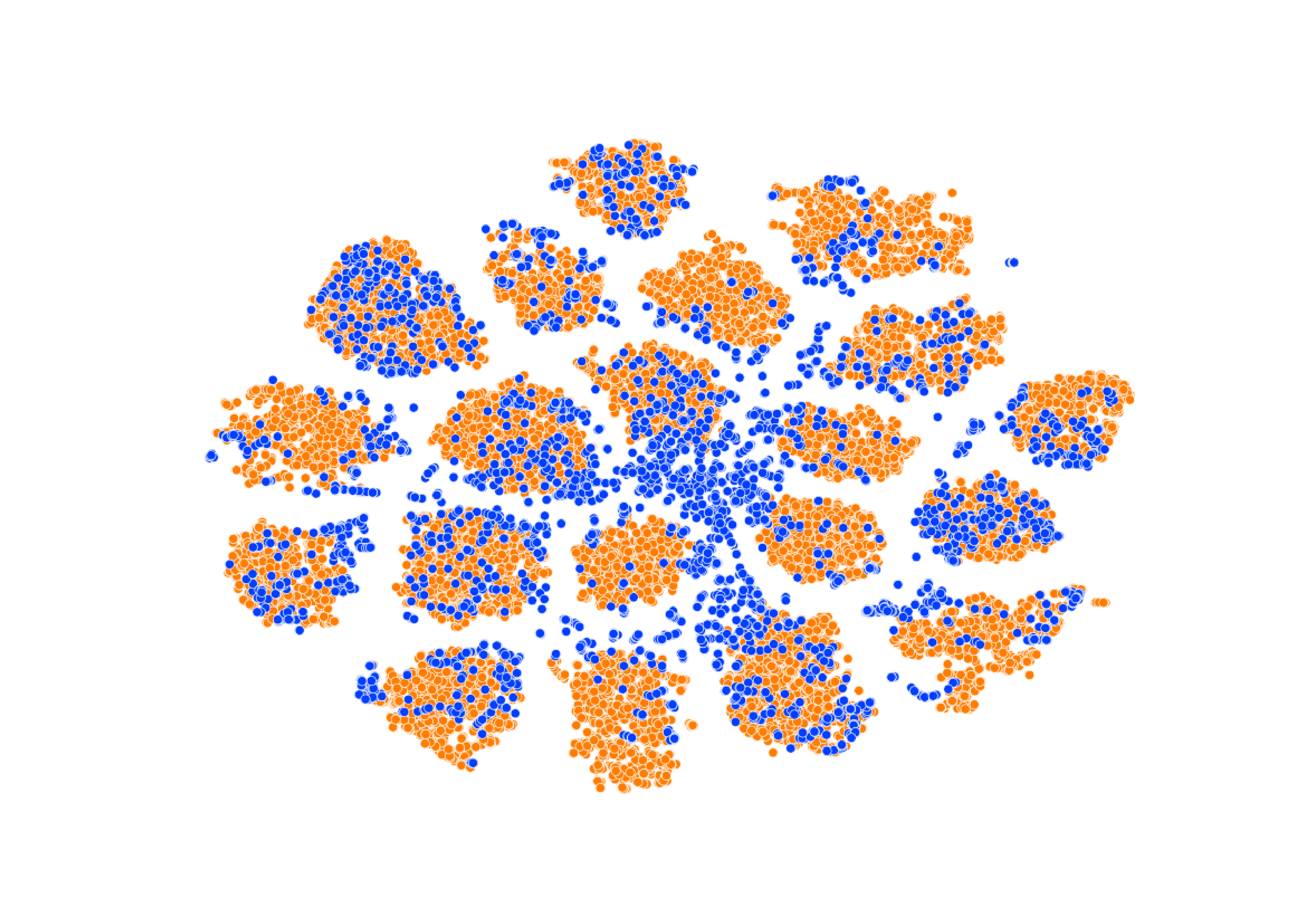}
\label{fig:source and target}
}
\caption{Visualization with t-SNE: (a)source domain data, each singer with four albums data in the train set. (b)target domain data, each singer with one album data in the test set. (c)the feature represents domain adaption, 0 represents the target domain data and 1 represents the source domain data.}
\end{figure*}

\subsection{Dataset and Audio Processing}
Artist20, a music artist classification dataset published by Ellis~\cite{ellis2007classifying}, was used in this paper to evaluate classification performance. 



The Artist20 dataset is the most widely used dataset for artist classification. It consists of 20 artists, each artist has 6 albums, and a total of 1413 songs. For the artist classification, there are different dataset split methods were used in related works. In addition to ensuring that there is no same frame was used both in the test song and train song, it is important to take into account the producer effect found by Whitman \textit{et al.}~\cite{whitman2001artist} in previous research. This refers to inflated classification performance in datasets split by song because of how salient production details can be in comparison to musical style. To combat this, the standard approach is to split the dataset by album such that the test set is composed solely of songs from albums not used in training. We divided four of the six albums into training sets, and the remaining two were divided into validation sets and test sets. We also do a preprocess on the data, cutting each song according to a window length of 30 seconds and a sliding time of 10 seconds~\cite{mcfee2015librosa,zhang2022MDCNN-SID}. The models are evaluated using Artist20 under the album split, averaging the F1 scores of three independent runs.

\subsection{Experimental Results}

In order to evaluate the performance of different algorithms, the metrics of F1 value are calculated based on precision and recall to evaluate them as a whole~\cite{li2008learning}. F1 score is the harmonic average of precision and recall, which can reflect the performance of the classification model more objectively.

Table \ref{tab:addlabel} shows the experimental results, we compared the proposed three-domain adaptation-based methods with the baseline method of CRNN proposed by Nasrullah~\cite{nasrullah2019music}. The four models share the same CRNN architecture of four CNN blocks and two GRU blocks. And the three proposed methods start with the pre-trained model weights of the CRNN.

\begin{table}[htbp]
  \centering
  \caption{Average testing F1 score on the Artist20 dataset}
    \begin{tabular}{lcc}
    \toprule
    Model & F1/best & F1/avg \\
    \midrule
    CRNN~\cite{nasrullah2019music}  & 0.61 & 0.60  \\
    CRNN-MMD &  0.76 & 0.74    \\
    CRNN-RevGrad &   0.82 & 0.81   \\
    \textbf{CRNN-CAN} &   \textbf{0.85} & \textbf{0.83} \\
    \bottomrule
    \end{tabular}%
  \label{tab:addlabel}%
\end{table}%

From the comparison results, the baseline method CRNN is state of the art, while it could achieve only 0.61 on the Artist20 dataset with album split. We use the CRNN-MMD can get an improvement of 0.15 than the baseline CRNN. The improvement of the proposed three methods shows the validity of the domain adaption. 

The CRNN-RevGrad and the CRNN-CAN get 0.82 and 0.85 on the F1 score separately. CRNN-RevGrad compare with the CRNN-MMD shows that a unified and robust feature representation between the source domain and target domain is more effective than simply minimizing the distance between domains. 

The CRNN-RevGrad through the weight splitting considers the independent source and target mappings for the feature representation. Due to the parameters being from the pre-trained model of the source domain data, it is more flexible for the domain-specific feature extractions. 

Finally, the CRNN-CAN compares with the CRNN-RevGrad that adding the correspondence between categories in different domains enhances the precise expression ability of domain adaptation.

\subsection{Visualization}
The last dense layer of the network processes a fully connected layer. And the audio sample is converted to a vector by the last dense layer. These vectors represented the feature learned by the network to distinguish the singer's label. Using t-Distributed Stochastic Neighbor Embedding (t-SNE) can further reduce the dimensional to visualize the classification by the learned feature. Figure \ref{fig:source domain} is the result of the training set with CRNN-CAN. It shows that on the source domain the sample is well separated into different clusters. While Figure \ref{fig:target domain} is the result on the target domain in which are different albums song of the test set. The learned feature representation can be distinguished well in the source domain but is not effective in the target domain. Figure \ref{fig:source and target} is the t-SNE result diagram of the CRNN-CAN model. The yellow dots in the figure are the source domain mapping diagrams, and the blue dots are the target domain mapping diagrams. It can be seen from the figure that the two domains have basically overlapped, and there is still room for improvement.



\section{Conclusions}

In this paper, we introduced the domain adaption method to address the album effect on singer identification task for the coming Metaverse. Three different methods were tried and combing with the backbone network of CRNN. The CRNN-MMD adds a dense layer to calculate the domain loss for the domain adaption. The CRNN-RevGrad adds a gradient reversal layer to try to learn a feature representation effective for the both source domain and target domain. While the CRNN-CAN adds the class information of the target domain by a clustering process between source data and target data. Experiments on the public dataset of Artist20 show that domain adaption can achieve a better result than the baseline method of CRNN. Finally, the proposed method of CRNN-CAN gets 0.85 in terms of F1 measure with the album split on Artist20. 
\section{Acknowledgement}
This paper is supported by the Key Research and Development Program of Guangdong Province under grant No.2021B0101400003. Corresponding author is Jianzong Wang from Ping An Technology (Shenzhen) Co., Ltd (jzwang@188.com).
\bibliographystyle{IEEEtran}
\bibliography{mybib}

\end{document}